\documentclass[
 aip,
 apl,
 amsmath,
 amssymb,
 reprint,
]{revtex4-1}

\usepackage[ansinew]{inputenc} 
\usepackage{graphicx}
\usepackage{SIunits}
\pdfoutput=1

\begin{document}

\title{Laser-sintered thin films of doped SiGe nanoparticles} 

\author{B. Stoib}
\email[Electronic mail: ]{benedikt.stoib@wsi.tum.de}
\author{T. Langmann}
\author{S. Matich}
\author{T. Antesberger}
\affiliation{Walter Schottky Institut, Technische Universität München, Am Coulombwall 4, 85748 Garching, Germany}
\author{N. Stein}
\affiliation{Faculty of Engineering and Center for Nanointegration (CeNIDE), University of Duisburg-Essen, Bismarckstraße 81, 47048 Duisburg, Germany}
\author{S. Angst}
\affiliation{Faculty of Physics and CeNIDE, University of Duisburg-Essen, Lotharstr. 1, 47048 Duisburg, Germany}
\author{N. Petermann}
\affiliation{Institut für Verbrennung und Gasdynamik and CeNIDE, University of Duisburg-Essen, Lotharstr. 1, 47048 Duisburg, Germany}
\author{R. Schmechel}
\author{G. Schierning}
\affiliation{Faculty of Engineering and Center for Nanointegration (CeNIDE), University of Duisburg-Essen, Bismarckstraße 81, 47048 Duisburg, Germany}
\author{D. E. Wolf}
\affiliation{Faculty of Physics and CeNIDE, University of Duisburg-Essen, Lotharstr. 1, 47048 Duisburg, Germany}
\author{H. Wiggers}
\affiliation{Institut für Verbrennung und Gasdynamik and CeNIDE, University of Duisburg-Essen, Lotharstr. 1, 47048 Duisburg, Germany}
\author{M. Stutzmann}
\author{M. S. Brandt}
\affiliation{Walter Schottky Institut, Technische Universität München, Am Coulombwall 4, 85748 Garching, Germany}


\begin{abstract}
We present a study of the morphology and the thermoelectric properties of short-pulse laser-sintered (LS) nanoparticle (NP) thin films, consisting of $\rm{SiGe}$ alloy NPs or composites of $\rm{Si}$ and $\rm{Ge}$ NPs. Laser-sintering of spin-coated NP films in vacuum results in a macroporous percolating network with a typical thickness of \unit{300}{\nano\meter}. The Seebeck coefficient is independent of the sintering process and typical for degenerate doping. The electrical conductivity of LS films rises with increasing temperature, best described by a power-law and influenced by two-dimensional percolation effects.
\end{abstract}

\pacs{68.55-a, 73.50.Lw, 73.63.Fg, 81.05.Rm, 81.07.Bc}
\maketitle 

Nanostructured thermoelectric materials have attracted substantial interest due to the prospect of an increased power factor, a reduced thermal conductivity $\kappa$ and the resulting enhancement of the energy conversion efficiency\cite{Dresselhaus2007,Snyder2008,Minnich2009,Kanatzidis2010}. On the one hand, a reduction in $\kappa$ can be obtained by classical alloying and the introduction of larger mass fluctuations such as inclusions or precipitates and crystallographic discontinuities as typically found in nanocrystalline materials. Also porosity\cite{Song2004,Tang2010} and surface modification\cite{Boukai2008} can reduce $\kappa$. The variety of length scales present should best be tuned to efficiently affect the propagation of the whole spectrum of phonons while, at the same time, maintaining good electronic properties. On the other hand, an enhancement of the power factor\cite{Heremans2011,Zebarjadi2011} can be achieved by a reduction of dimensionality, bandgap engineering, modulation doping or strain. Classical thermoelectric materials such as SiGe\cite{Wood1988,Slack1991} profit from nanostructuring\cite{Zhu2009} as well as materials such as pure Si, which has recently been reconsidered for thermoelectric applications\cite{Hochbaum2008,Boukai2008,Tang2010,Petermann2011}. Besides their technical relevance, group-IV semiconductors with their well known properties are an ideal model system to study the benefits of nanostructures on the overall thermoelectric performance. In particular thin films are highly suited for such fundamental investigations since they allow access to many analytical methods and provide a range of potential parameters for optimization such as the compatibility factor in segmented thermoelectrics\cite{Seifert2010}. In this work we report on laser-sintered (LS) thin layers of $\rm{SiGe}$ nanoparticles (NPs), the morphology of the thin films obtained and the influence of porosity on the homogeneity of electrical conduction. Both the Seebeck coefficient~$S$ and the temperature-dependent macroscopic electrical conductivity $\sigma$  are determined and compared to bulk samples produced by current-assisted sintering (CAS) of the same initial NPs\cite{Stein2011}. 

The NPs were synthesized by plasma-assisted decomposition\cite{Knipping2004} of silane and germane. Decomposition of both gases in a single microwave plasma leads to a homogeneous mixing of $\rm{Si}$ and $\rm{Ge}$ within each NP\cite{Stein2011} (henceforth called alloy NPs, denoted $\rm{Si}_{x}\rm{Ge}_{y}$). Separate growth of $\rm{Si}$ NPs and $\rm{Ge}$ NPs in two microwave plasmas and their subsequent mixing in the gas phase leads to a homogeneous mixing of the two sorts of NPs (henceforth called NP composites, denoted $\rm{Si}\text{-}\rm{Ge}\,\text{x:y}$). The Si, Ge as well as alloy NPs obtained are highly crystalline and spherical with a mean diameter of \unit{20\text{-}30}{\nano\meter}. For all samples in this work, 1\,\% per volume of phosphine was admixed to the silane source as dopant. A native oxide shell is formed due to an unavoidable exposure to air upon removal of the NPs from the plasma reactor. Previous studies on pure $\rm{Si}$ NPs have shown that the overall $\rm{P}$ incorporation efficiency is almost 100\,\%. However, about 95\,\% of the dopants accumulate in the outer shell of the NP which is oxidized upon exposure to air, while only 5\,\% remain within the core and are electrically active\cite{Stegner2009}. An analogous secondary ion mass spectroscopy of the $\rm{Si}_{80}\rm{Ge}_{20}$ NP films studied here suggests that P accumulation at the NP surface is reduced in alloy NPs, with approximately 40\,\% of the $\rm{P}$ atoms incorporated in the un-oxidized core.

The NP powders were dispersed in ethanol (5\,\% by weight). Spin coating at \unit{2000}{rpm} on flexible polyimid foils (Kapton\textsuperscript{\textregistered} HN \unit{125}{\micro\meter}) resulted in homogeneous layers of typically \unit{300}{\nano\meter} thickness. Etching in an aqueous solution of 5\,\% hydrofluoric acid (HF) for \unit{2}{min}, rinsing in deionized water and flushing with dry $\rm{N}_2$ efficiently removed the native oxide. The films were transferred to a vacuum chamber with a borosilicate window within \unit{5}{min}. Laser-sintering was performed at a base pressure of less than $\unit{5\cdot 10^{-5}}{mbar}$ with a frequency-doubled $Q$-switched Nd:YAG laser at $\lambda=\unit{532}{\nano\meter}$, a pulse length of \unit{5\text{-}7}{\nano\second} and a repetition rate of \unit{10}{Hz}. The pulse energy could be adjusted up to \unit{650}{\milli\joule} for a beam diameter of \unit{8}{\milli\metre}. For all NP layers studied in this work an average energy density of \unit{100\text{-}120}{\milli\joule\per\centi\metre^2} led to the best results in terms of $\sigma$. The sintering was performed at constant energy density, moving the sample under the beam at a constant speed of \unit{0.5}{\milli\metre\per\second}. The advantages of this ``moving spot'' approach are twofold: After HF etching the NPs are partly H-terminated\cite{Stegner2008} and residual solvent is contained inside the macroporous film. The overall Gaussian beam shape in combination with the movement of the sample through the beam leads to a pre-heating which drives out hydrogen and solvents which otherwise would have left the film explosive, leading to ruptures in the morphology. Furthermore, this approach averages out inhomogeneities in the laser beam which cannot be removed optically. 

Secondary electron micrographs (SEM) were obtained using a Hitachi S3200N. Scanning transmission electron micrographs (STEM) were obtained using a Zeiss NVision~40 equipped with a segmented annular as well as an on-axis detector. Since the minimum angle for the annular detector was \unit{10}{mrad}, gray-levels cannot only be ascribed to a compositional contrast\cite{Pennycook1989}. For thermoelectrical measurements silver contacts were sputtered onto the samples. Temperature-dependent conductivity measurements were preformed in two-point coplanar geometry in vacuum applying a ramp speed of \unit{1}{\kelvin\per\min}. To minimize the influence of adsorbate desorption on $\sigma$ we only show cooling curves. $I\text{-}V$ characteristics demonstrated ohmic behaviour at all temperatures. Profilometry was used to determine the average macroscopic height $h$ of the films with width $w$ and length $l$. Only these quantities were used to calculate a macroscopic $\sigma$, not accounting for porosity. Seebeck measurements were performed using a direct method applying two type\,K thermocouples\cite{Brandt1998}. The thermovoltage $U$ was picked up between the alumel legs and corrected for $S_{alumel}=\unit{-19}{\micro\volt\per\kelvin}$ for the whole temperature range investigated here\cite{Bernhard2004}. $S$ was deduced from the slope of $\Delta U$ vs. $\Delta T$, maintaining a constant mean temperature. For all measurements $\Delta T$ was kept smaller than \unit{10}{\kelvin}. Bulk CAS samples produced of the same initial NP powders as LS samples were measured with a commercial setup (ZEM-3 by Ulvac Technologies, Inc.) as well as with the apparatus just described to check for the consistency of the measurement setups.

\begin{figure}	
	\includegraphics[width=8.5cm]{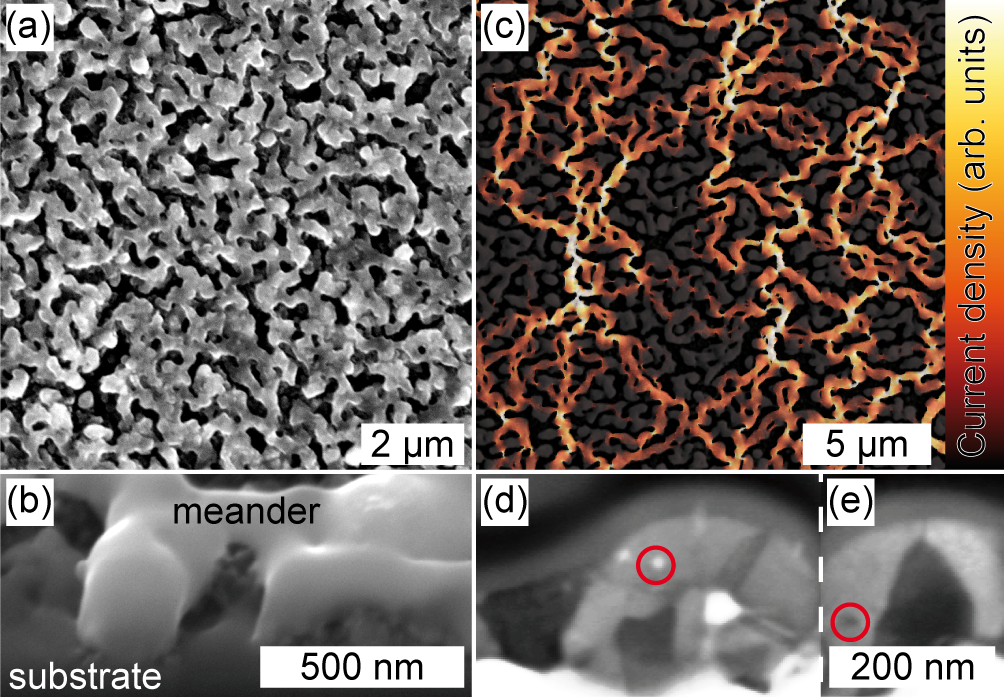}
	\caption{(a)~Secondary electron microscopy (SEM) top view of a laser-sintered (LS) $\rm{Si}\text{-}\rm{Ge}~\text{91:09}$ thin film showing the porosity of the meander structure typical for all samples investigated in this work. (b)~Side view of the LS film illustrating the partial substrate coverage. (c)~Superposition of the SEM image of a LS film of $\rm{Si}_{80}\rm{Ge}_{20}$ NPs as the basis of a resistor network simulation and a false color image of the current density, as obtained by this simulation. The electric field is applied between top and bottom of the image. (d)~and~(e)~Cross-sectional scanning transmission electron micrographs (STEM) of a LS $\rm{Si}\text{-}\rm{Ge}~\text{60:40}$ composite thin film. Circles indicate inclusions.}
	\label{fig:morphology}
\end{figure}
Figure~\ref{fig:morphology}(a) shows a top view SEM image of the general sample morphology of LS thin films which forms connected, meander-like structures constituting a macroporous network covering the substrate to typically 60-80\,\%\footnote{LS under ambient conditions led to rather dewetted isolated drops. No sufficient electrical conduction could be established.}. The meander structure is similar for alloy and composite films studied here. The side view in Fig.~\ref{fig:morphology}(b) shows the partial coverage of the substrate by the meander network as well as its varying height of $\unit{200\text{-}300}{\nano\metre}$, which is a characteristic length of the macroporous meander structure. The side view suggests that in-plane electrical transport is governed by an effectively two-dimensional network, with a percolation threshold $p_{th}$ of $\approx 50\text{-}60\,\%$ (depending on the assumed lattice) and a slow increase of conductivity above $p_{th}$\cite{Last1971}. To study the effect of substrate coverage and porosity on the conductance we applied a random resistor network simulation\cite{Knudsen2006} on a square lattice where we exemplarily converted a series of typical SEM pixel images into networks of equal resistors. A pixel belonging to the meander layer is assigned to be conductive while a pixel belonging to pore space is nonconductive. Assuming a certain external current entering the resulting resistor network on one side and leaving on the opposite side, the local potentials are calculated. The local current densities are then obtained from the corresponding potential differences. Figure~\ref{fig:morphology}(c) shows that electrical conduction is mainly carried by a few percolating conduction paths. In the example shown, the macroscopic resistivity is increased by a factor of 14 compared to a fully conductive layer without meanders and holes. Although alloys and composites form similar meander structures on the \unit{\micro\meter} scale and have similar grain sizes of $\approx\unit{100\text{-}150}{\nano\meter}$ (see Fig.~\ref{fig:morphology}(d) and (e)) they differ on a tens of \unit{\nano\meter} scale. Whereas LS alloys form smooth meanders, LS composites show grainy features of \unit{20\text{-}50}{\nano\meter} size on the surface as well as within grains (circles). We ascribe this to inclusions of mostly as-grown NPs, unaffected by LS. 

\begin{figure}	
	\includegraphics{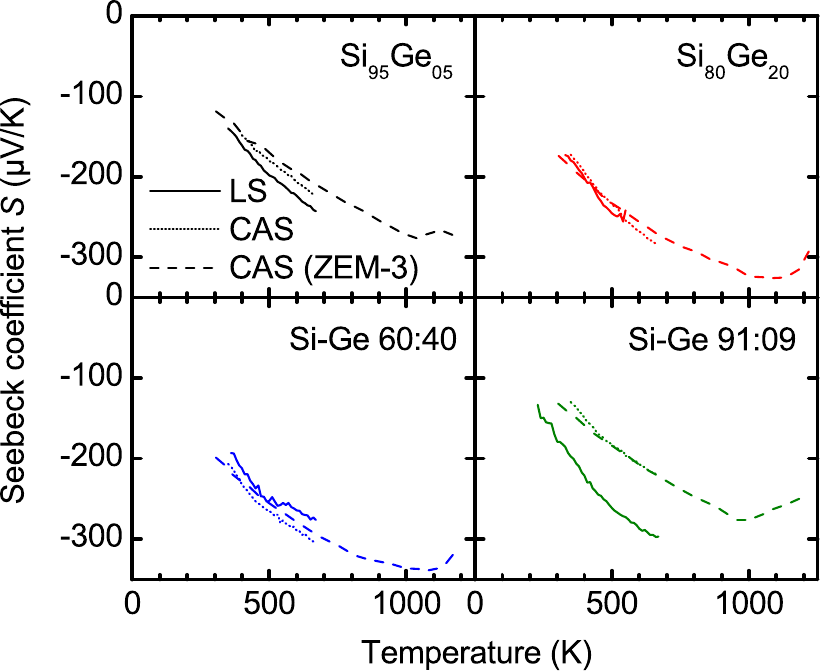}
	\caption{Seebeck coefficient $S$ of samples of four $\rm{SiGe}$ mixtures prepared by laser-sintering (LS) and current-assisted sintering (CAS). $S$ is largely independent of the sintering process, showing the behaviour of degenerately doped $\rm{SiGe}$.}
	\label{fig:seebeck}
\end{figure}
In Fig.~\ref{fig:seebeck} the Seebeck coefficient of LS $\rm{Si}_{80}\rm{Ge}_{20}$ and $\rm{Si}_{95}\rm{Ge}_{05}$ alloys and LS $\rm{Si}\text{-}\rm{Ge}~\text{91:09}$ and $\rm{Si}\text{-}\rm{Ge}~\text{60:40}$ composites is shown together with the results of the corresponding CAS samples of the same initial raw materials. The results obtained by the two Seebeck set\-ups agree quantitatively, demonstrating the high reproducibility of the measurments. All samples show a negative $S$ whose absolute values increase with increasing temperature. This can be understood due to the heavy $\rm{P}$ doping above the metal-insulator transition\cite{Dismukes1964,Conwell1956}. For all samples the measured data can well be extrapolated to $S=\unit{0}{\micro\volt\per\kelvin}$ at $T=\unit{0}{\kelvin}$, as expected for degenerate doping. Taking into account the differences in the microstructure of CAS and LS samples, it is encouraging to see that the Seebeck coefficients $S$ of alloys and composites treated by these two sintering methods agree well. Assuming Si-like effective masses\cite{Green1990} $m^{*}$, the n-type carrier concentration $n$ can be evaluated according to Ref.~\citenum{Snyder2008} via $S=\frac{8 \pi^2 k_{B}^2}{3 e h^2} m^{*} T (\frac{\pi}{3n})^{2/3}$, with the Boltzmann constant $k_{B}$, the Planck constant $h$ and the ele\-mental charge $e$. For all LS samples $n\approx \unit{4 \text{-} 9 \cdot 10^{19}}{\centi\meter^{-3}}$.

\begin{figure}	
	\includegraphics{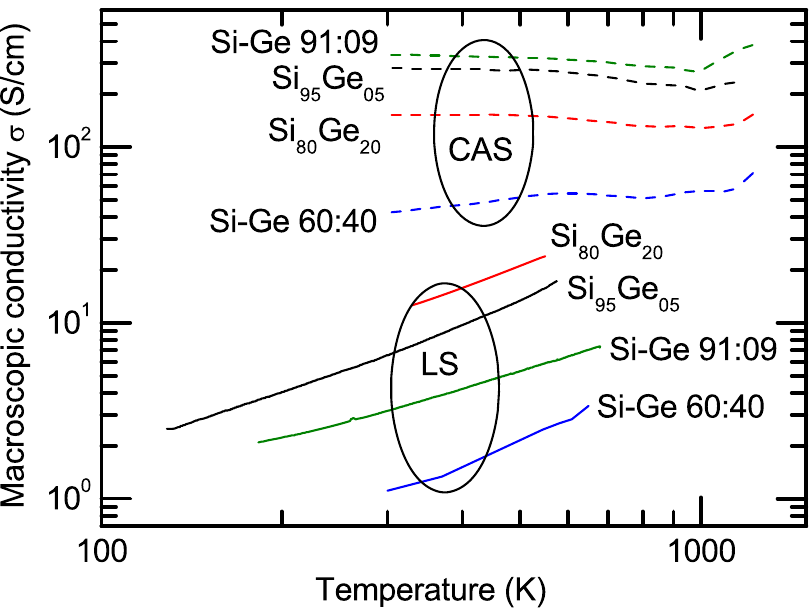}
	\caption{Comparison of the macroscopic electrical conductivity $\sigma$ of LS thin films and CAS samples. Data of LS films were not corrected for porosity. While CAS samples show classical metallic behaviour, the LS films exhibit a power-law $\sigma \propto T^{\sim 1.2}$.}
	\label{fig:conductivity}
\end{figure}
Figure~\ref{fig:conductivity} shows the $T$-dependence of the macroscopic electrical conductivity $\sigma$ of LS and CAS samples. In the case of degenerate doping the number of charge carriers should be temperature-independent with the conductivity limited by the mobility. While CAS samples show a flat behaviour, the LS samples show an increasing conductivity with temperature. Furthermore, LS samples differ from CAS samples in the absolute value of $\sigma$, which can at least partly be ascribed to percolation effects in LS samples, discussed in conjunction with Fig.~\ref{fig:morphology}(c), which were not corrected in Fig.~\ref{fig:conductivity}. Empirically, the observed temperature dependence of $\sigma$ for LS samples can be described by a power-law with $\sigma \propto T^{\sim 1.2}$. 

\begin{figure}	
	\includegraphics{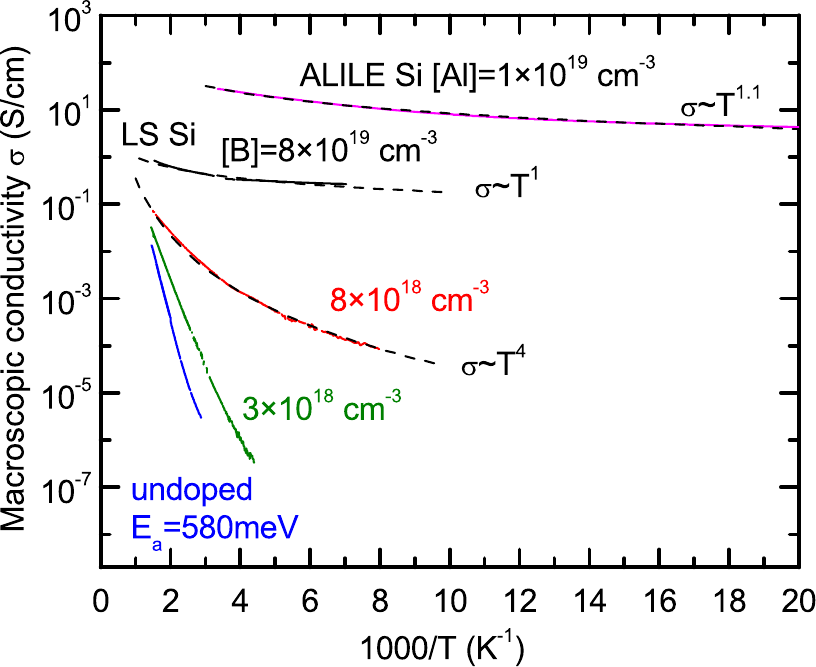}
	\caption{Study of the influence of B doping in macroporous laser-sintered (LS) Si films. Dashed lines indicate power-laws. The conductivity obtained on a continuous poly-Si film fabricated via aluminum-induced layer exchange (ALILE) is shown as a reference.}
	\label{fig:siliconboron}
\end{figure}

To discuss this finding we additionally present a more systematic study of the influence of doping in a simplified system, namely LS films of pure Si NPs. Due to availability, those films were doped with B, which should not alter the principal findings. A doping series of LS Si NP films is shown in Fig.~\ref{fig:siliconboron}. Presented in an Arrhenius plot power-law dependencies $\sigma\propto T^\alpha$ appear bent (dashed lines). Experimental data of samples with different doping levels ranging from $\unit{3 \cdot 10^{18}}{\centi\meter^{-3}}$ to $\unit{8 \cdot 10^{19}}{\centi\meter^{-3}}$ show a correlation of doping level and the apparent exponent $\alpha$. Typical for undoped Si, a nominally undoped film shows an activated exponential behaviour with $E_{a}\approx\unit{580}{\milli e \volt}$. This exponential behaviour is also observed for the least doped film at higher temperatures, indicating a continous change from power-law to exponentially activated behaviour. Since LS Si NP films have a similar morphology as the samples discussed above, percolation also affects their absolute value of $\sigma$. As a polycrystalline reference without porosity but with comparable p-type doping, a sample prepared by aluminum-induced layer exchange\cite{Antesberger2008} (ALILE) is shown. The $T$-dependence of $\sigma$ is similar to LS samples but the absolute values of $\sigma$ are higher for the ALILE sample, confirming the influence of percolation on LS samples. 

Our model to explain the temperature dependence is that the transport is governed by an interplay of doping density within the grains and a distribution of barrier heights across the grains. For high doping and low temperature, a certain subensemble of the lowest barriers can be regarded as conductive due to screening by free charge carriers. With increasing temperature more conductive paths are available because also grain boundaries with higher barriers become transparent for charge carriers. We believe that this increased number of possible paths is responsible for the bending in an Arrhenius plot. We note that in recent years a power-law of the electrical conductivity was reported in a number nano-scale systems (see, e.\,g., Ref.~\citenum{Rodin2010} and references therein). 

Reducing $\kappa$ via nanostructuring while maintaining/enhancing the power factor are mostly contradicting goals. Introducing disorder via alloying, fine graining, inclusions and porosity easily will lower $\kappa$. We showed that the bottom-up approach of laser-sintering SiGe nanoparticles leads to promising morphological properties, with alloy disorder on the atomic scale and structural features with sizes in the \unit{20\text{-}50}{\nano\meter} (inclusions), \unit{100\text{-}150}{\nano\meter} (grains) and \unit{200\text{-}300}{\nano\meter} (macroporosity) ranges. The fact that the Seebeck coefficient is in agreement with other SiGe systems with similar doping and that the electrical conductivity rises with temperature leads to the conclusions that LS samples already show high quality within the grains and that transport across grain boundaries is largely affected by potential barriers with a wide distribution of barrier heights, depending on the local defect and doping density. The discrepancy to CAS samples suggests a different chemical environment of the grain boundary. Thus, future experiments including H-passivation, furnace or microwave post-treatments will be helpful to understand the transport mechanism and to find strategies for further improvement. Furthermore, overcoming the limitation of two-dimensional percolation can be achieved by filling the remaining pores with a second layer. First studies already showed that the layer deposition sequence of spin-coating, etching and laser-sintering can be applied several times. The meander-like structures of different layers produced are partially sintered together, increasing the degree of interconnection but still maintaining a high degree of porosity.

This work was supported by Deutsche Forschungsgemeinschaft via priority program SPP~1386.


%

\end{document}